\documentclass[aps,prx,twocolumn,superscriptaddress,floatfix]{revtex4-2}
\usepackage{graphicx} 
\usepackage{amsmath, amssymb, braket, dsfont}
\usepackage{xcolor, comment}

\date{\today}
\begin{document}
\title{Classical benchmarking of zero noise extrapolation beyond the exactly-verifiable regime}

\author{Sajant Anand}
\affiliation{Dept. of Physics, University of California, Berkeley, CA 94720, USA}
\author{Kristan Temme}
\affiliation{IBM Quantum, IBM Thomas J. Watson Research Center, Yorktown Heights, NY, USA}
\author{Abhinav Kandala}
\affiliation{IBM Quantum, IBM Thomas J. Watson Research Center, Yorktown Heights, NY, USA}
\author{Michael Zaletel}
\affiliation{Dept. of Physics, University of California, Berkeley, CA 94720, USA}
\affiliation{Material Science Division, Lawrence Berkeley National Laboratory, Berkeley, California 94720, USA}

\begin{abstract}

    In a recent work a quantum error mitigation protocol was applied to the expectation values obtained from circuits on the IBM Eagle quantum processor with up $127$ - qubits with up to $60 \; - \; \mbox{CNOT}$ layers. To benchmark the efficacy of this quantum protocol a physically motivated quantum circuit family was considered that allowed access to exact solutions in different regimes.
    The family interpolated between Clifford circuits and was additionally evaluated at low depth where exact validation is  practical. It was observed that for highly  entangling parameter regimes the circuits are beyond the validation of matrix product state and isometric tensor network state approximation methods. Here we compare the experimental results to matrix product operator simulations of the Heisenberg evolution, find they provide a  closer approximation than these pure-state methods by exploiting the closeness to Clifford circuits and limited operator growth. Recently other approximation methods have been used to simulate the full circuit up to its largest extent. We observe a discrepancy of up to $20\%$ among the different classical approaches so far, an uncertainty comparable to the bootstrapped error bars of the experiment. Based on the different approximation schemes we propose modifications to the original circuit family that challenge the particular classical methods discussed here.

\end{abstract}

\maketitle

Quantum error mitigation (QEM) has been proposed as a method for extending the reach of near-term quantum hardware before the implementation of quantum error correction~\cite{Temme2017Error,Li2017Efficient}. While quantum error correction is widely expected to be necessary for implementing generic quantum algorithms, the overhead is currently prohibitive. QEM instead compensates for the effect of errors by engineering the noise in a manner which allows postprocessing to increase the accuracy of expectation values. 

\begin{figure}[bt]
    \centering
    \includegraphics{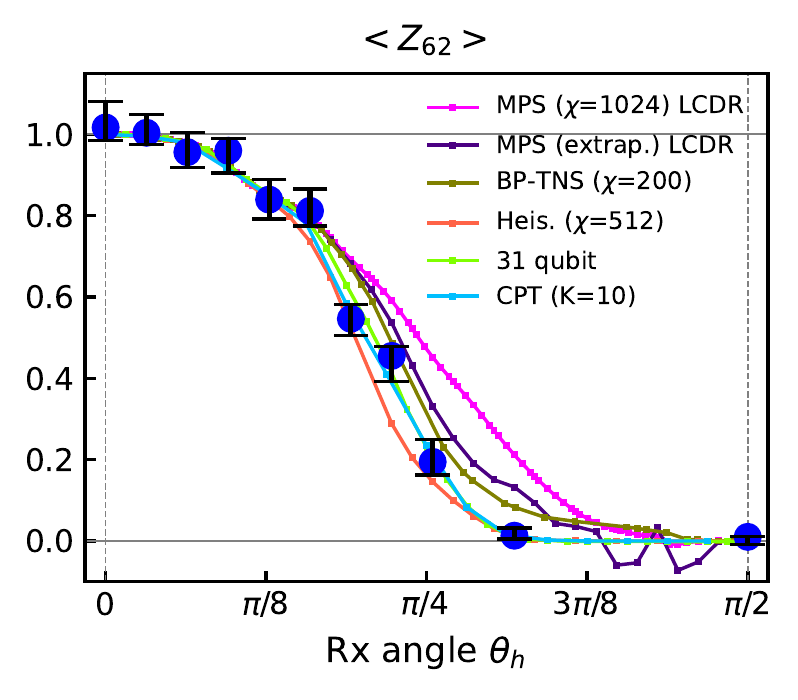}
    \caption{Comparison of classical approximations for $\langle Z_{62}\rangle$ at Trotter depth $D=20$ against experimental ZNE results: (1) matrix-product-state  (MPS) representation of the pure state within a lightcone-reduced volume \cite{utility}; (2) extrapolation of the MPS results with respect to the estimated circuit fidelity (see Supp.); (3) Belief propagation tensor network states~\cite{tindall2023efficient}; (4) MPO representation of Heisenberg evolution (this work); (5) simulation of a 31 qubit subset of the IBM Eagle device~\cite{kechedzhi2023effective}; (6) Clifford perturbation theory~\cite{begušić2023fast}. The latter four methods differ by $\sim 20\%$ amongst themselves near $\theta_h \sim \pi/4$, an amount largely within the spread of the ZNE error bars. Without further calculations it is not clear which of these methods is most accurate.  }
    \label{fig:Z62}
\end{figure}

In a recent work~\cite{utility}, we reported results benchmarking the efficacy of  an experimentally feasible QEM variant,  ``zero noise extrapolation'' (ZNE)~\cite{Temme2017Error,Li2017Efficient,Kandala2019Error} for estimating observables using error-prone quantum hardware. To do so, we considered a quantum circuit describing the discretized dynamics of the 2D transverse field Ising model (TFI)  (i.e., the ``kicked'' transverse Ising model \cite{prosen2000exact,pineda2014two}).
Each round of the circuit takes the form
\begin{align}
U(\theta_J, \theta_h) &= \prod_{\braket{i,j}}R_{ZZ}^{ij}(\theta_J) \prod_{i} R_X^i(\theta_h) \nonumber \\
&= \prod_{\braket{i,j}}e^{-i \theta_J Z_i Z_j / 2} \prod_{i}e^{-i \theta_h {X_i}/2} 
\label{eq:utilityTFI}
\end{align}
Here $\langle i, j \rangle$ runs over the bonds of IBM's ``Eagle’’ 127-qubit heavy-hexagon lattice.  
The two-qubit angle $\theta_J=-\pi/2$ was chosen to be Clifford so as to involve only a single CNOT gate, and we assume this value unless otherwise specified.
At $\theta_h = 0, \pi/2$ the overall circuit dynamics are Clifford and therefore exactly calculable, while at intermediate $\theta_h$ they are expected to be ergodic. 

In the reported experimental protocol, the system is first initialized in  state $\otimes^{127}_{i=1} \ket{\uparrow}$ and evolved under $D$-rounds of $U$, at which point a variety of weight-1, 10 and 17 observables were estimated. Results  were compared with exact classical calculations where available, as well as 1D and 2D tensor network simulations which approximate the evolution of the pure quantum state. These classical approximations are expected to break down at large depths due to the growth of quantum entanglement, which is particularly rapid at $\theta_h = \pi /2$.

Experimentally, ZNE was found to provide a large improvement in the accuracy of expectation values compared with unmitigated results. We found:
\begin{enumerate}
\item{} In the regime where exact classical simulations can verify results (including circuits with a depth of 15 CNOT layers, in which the measured operator spread up to 68 sites), ZNE reproduces the exact results within the bootstrap error bars (Fig.~\ref{fig:Z62})

\item{} Classical matrix-product state (MPS) and 2D isometric tensor-network (isoTNS) methods which approximate the dynamics of the pure state, while accurate near $\theta_h \sim 0$, struggle to accurately reproduce expectation values for the full range of $\theta_h$ values. 

\item{} In regimes beyond exact verification, ZNE produces results which are correct at both of the exactly solvable points $\theta_h = 0, \pi/2$,  while the classical pure-state approaches fail  badly near $\theta_h \sim \pi/2$. 
\end{enumerate}

The observation that ZNE could provide such reliable expectation values, at this scale, was argued to be evidence for the utility of pre - fault tolerant quantum computers. The argument for utility is that one can use noisy quantum processors as a tool to reliably explore circuits and problems that will ultimately pose difficult challenges for classical simulation methods. It is therefore crucial to emphasize that the Ref.~\cite{utility} did not claim “quantum advantage” or any provable speedup, and it was left open whether other approximate classical methods might perform better than the pure state methods. Indeed, it was suggested that approximating the Heisenberg evolution of the measured operators, rather than the quantum states themselves, was a promising future direction.

In this work we report the results of such additional classical calculations.
In contrast to the pure-state methods, we find that in the verifiable regime, matrix-product approximations of the Heisenberg evolution give excellent agreement with the exact results, and therefor with ZNE. 
Going beyond the verifiable regime, we find classical simulations remain in good agreement with ZNE even up to 20 Trotter steps (60 CNOT layers), and reproduce the exact results at $\theta_h = \pi/2$. 
These results provide further evidence that ZNE produces accurate expectation values at scales orders of magnitude beyond both exact verification and the unmitigated results.

At depth $D=20$, there remains a range of $\theta_h \sim \pi / 4$ where Heisenberg simulations are not fully converged, and they do not precisely agree with results of the newly developed ``BP-TNS'' classical approximation recently reported in~\cite{tindall2023efficient}; Clifford perturbation theory (CPT) reported in ~\cite{begušić2023simulating,begušić2023fast}; and simulations on a smaller 31-qubit geometry reported in~\cite{kechedzhi2023effective}. The various methods mutually disagree at about the 20\% level, which happens to be comparable to the ZNE error bars which sit between them - see  Fig.\ref{fig:Z62}.
The Heisenberg evolution is a fully controlled approximation, meaning it is in principle exact, but only in the limit of a bond dimension which scales exponentially in $D^2$. BP-TNS, on the other hand, is fully converged at a more favorable bond dimension of $2^D$, but makes uncontrolled approximations which make it inexact even in this limit. 
CPT is a perturbative expansion in $\tan(\theta_h)$ (truncated at order $K = 10$ in Ref.~\cite{begušić2023fast}) which is no longer a small parameter at the  $\theta_h \sim \pi / 4$ point.
Smaller-size methods depend on the restricted growth of the operator, and ultimately will require a careful finite-size scaling analysis. 
Without additional study, it is not yet clear which of the methods is most accurate in this regime.
We anticipate further improvements in classical methods or quantum hardware will prove fruitful here. 

\begin{figure}[tb]
    \centering
    \includegraphics{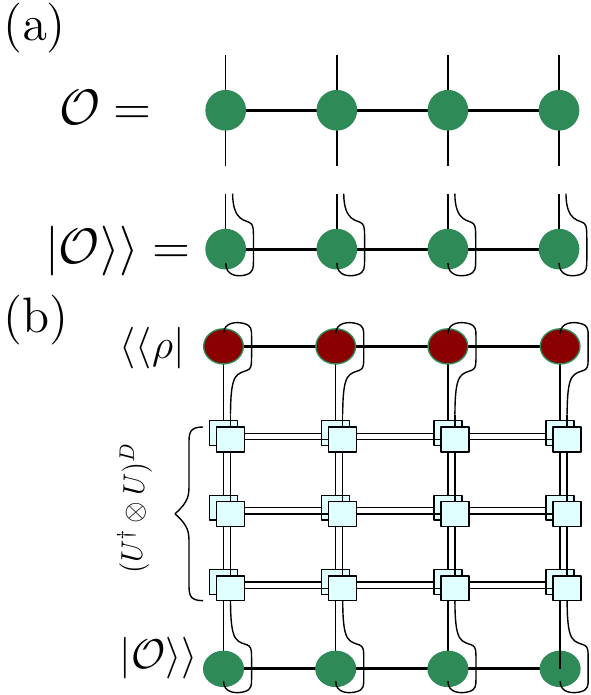}
    \caption{Evolution of operators using matrix product operators (MPO). (a) The 1D MPO representation of an operator can be viewed as a vectorized state in a larger Hilbert space of onsite dimension $k=4$. (b) An operator expectation value is found by evolving the vectorized operator by $U^\dagger \otimes U$, in this work represented by the product of 13 MPOs, each of bond dimension 4.}
    \label{fig:operatorMPO}
\end{figure}

\section{Heisenberg-MPO Method}
\label{sec:heisenberg}

\begin{figure*}[tb]
    \centering
    \includegraphics{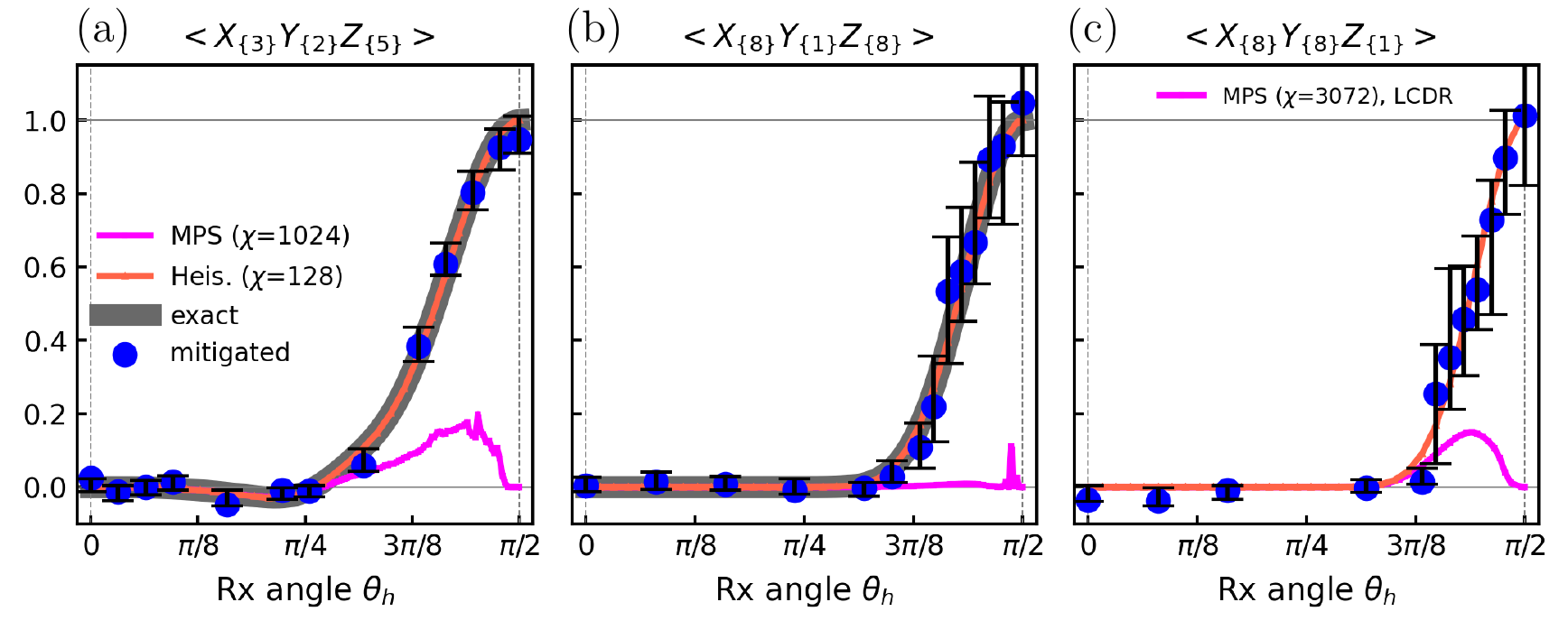}
    \caption{Comparison of Heisenberg evolution against experimental ZNE results for (a) weight-10, (b) weight-17, and (c) weight-17 with modified circuit, all at depth 5. For the first two, exact answer is available either by brute force or lightcone-reduced MPS simulations, and Heisenberg evolution accurately matches this. For (c), where the exact answer is not available, Heisenberg and ZNE agree.}
    \label{fig:hwZNE}
\end{figure*}

We begin by describing the numerical method used to approximate the Heisenberg evolution of an observable, $\mathcal{O}(D) = (U^\dagger)^D \, \mathcal{O} \, U^D$.  
We take the standard approach of approximating $\mathcal{O}(D)$ via a 1D matrix-product operator representation (MPO)~\cite{verstraete2004matrix, zwolak2004mixed}.
The MPO can be viewed as a vectorized MPS in a doubled Hilbert space of local bond dimension 4, as shown in Fig.~\ref{fig:operatorMPO}(a).
To map the 2D heavy-hex lattice to a 1D chain, sites are ordered according to the ``snake’’ ordering of Ref.~\cite{utility}.
This particular ordering minimizes the number of long-range connections in the resulting MPS.
As in the standard method for time-evolving 1D operators, we interleave unitary conjugation with variational matrix-product compression~\cite{schollwock2011density}.

However,  mapping  the 2D lattice to the 1D chain introduces long-range couplings which require some care to implement efficiently.
If a full round of $U$ is applied, the MPO dimension would increase by a factor of 4096, making subsequent truncation intractable\footnote{We note that the bra and ket legs of the operator must be evolved together, without truncation in between. If not, evolution of a Pauli string under a non-trivial Clifford circuit is not guaranteed to yield a single Pauli string.}.
Instead, we decompose $U$ into 13 layers, $U = \prod_{r = 1}^{13} U_r$, chosen such that the exact application $O \to U_r^\dagger O  U_r$ increases $\chi \to 4 \chi$ on the evolved bonds.
As the two-qubit gates commute, the layers $U_r$ are chosen to evenly distribute the gates among the layers.
To prevent blowup in $\chi$, two-site variational matrix product compression \cite{schollwock2011density} is then applied between each layer.
Simulations are conducted at fixed matrix-product bond dimension $\chi$, leading to errors at long times.
After $D$ steps, we then compute $\langle \mathcal{O}(D) \rangle = \braket{ \psi | \mathcal{O}(D)| \psi}$ for $\ket{\psi} = \otimes_{i=1}^{N} \ket{\uparrow}$, $N=127$.
Note that the evolved operator can be measured with any initial state, pure or otherwise.
This is shown in Fig.~\ref{fig:operatorMPO}(b) in the vectorized picture.
Each Trotter step takes time $N \chi^3$, for total complexity $D N \chi^3$. 

\section{Benchmarking the Kicked Ising Circuits}
\label{sec:utility}
We now consider the circuits of Ref.~\cite{utility}. In Fig.~\ref{fig:hwZNE}, we show results for the weight-10 operator $X_{\{3\}} Y_{\{2\}} Z_{\{5\}} = U(\pi/2)^5 Z_{13} \left[U^\dagger(\pi/2)\right]$ (a) and the weight-17 operator $ X_{\{8\}} Y_{\{1\}} Z_{\{8\}} = U(\pi/2)^5 Z_{58} \left[U^\dagger(\pi/2)\right]^5$ (b), both measured at circuit depth 5.
Both of these operators trivially have expectation value 1 (0) at the Clifford $\theta_h = \pi/2$ ($0$) point. 
We find that the results from Heisenberg evolution of the measured operator agree to good precision with the exact results, and thus with the ZNE-experimental results, across all ranges of $\theta_h$.
In Appendix~\ref{sec:numerics}, we quantitatively compare to the exact results and demonstrate agreement better than $10^{-4}$ (often significantly)  for all $\theta_h$.
In fact, for the weight-10 operator in Fig.~\ref{fig:hwZNE}(a), a bond dimension of $\chi=384$ is sufficient for exact simulation.
Note that unlike for pure state methods, which were considered previously in Ref.~\cite{utility}, the bond dimension needed at the non-trivial Clifford point $\theta_h=\pi/2$ is 1, despite the circuit generating a volume law entangled stabilizer state.
This is because a Pauli operator remains a single Pauli string, albeit expanded in extent, throughout the evolution.
Moving away from the either Clifford point $\theta_h = 0, \, \pi/2$ has a perturbative effect on the Heisenberg evolution, as the evolved operator becomes a superposition of Pauli strings concentrated in the vicinity of the Pauli string resulting from Clifford evolution.
Thus near either Clifford point, Heisenberg evolution performs well and needs minimal bond dimension to capture the dynamics.

\begin{figure*}[t]
    \centering
    \includegraphics{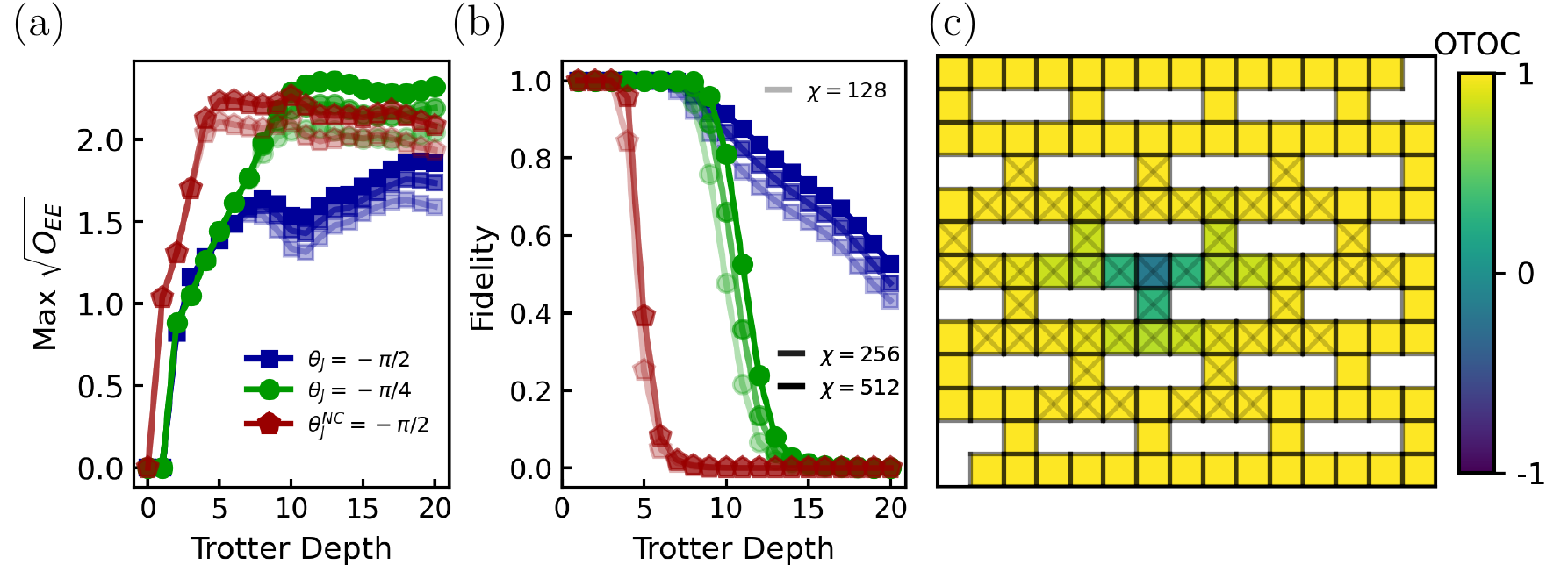}
    \caption{Operator spreading in the heavy-hex Kicked Ising dynamics. Growth in operator entanglement entropy (a) consistent with $O_{EE}(D) \propto D^2$ is consistent with an operator spreading over a disk. The plateau is an artifact of finite-$\chi$, which causes the circuit fidelity (b) to decay. Data is for the Kicked TFI with $\theta_J=-\pi/2$, a non-Clifford version $\theta_{J} = -\pi/4$, and modified $\theta_J=-\pi/2$ Kicked TFI model with additional layers of $R_X(\theta_h)$ gates between each of the three layers of $R_{ZZ}$ gates that constitute a Trotter step so that the two-qubit gate layers become non-commuting. All models have $\theta_h=0.7$. (c) Spatial profile of the out-of-time order correlator $C(7, x) = \langle Z_x Z_{62}(7) Z_x Z_{62}(7)\rangle $ for the experimental model, showing slow spread of the OTOC with circuit depth. The 54 sites in the lightcone at depth $D=7$ are shown by the crosses.}
    \label{fig:whyWorks}
\end{figure*}

Having built confidence in Heisenberg MPO evolution to match the exact results where available, we move to circuits where an exact result is not available.
To clarify, by exact we mean either a full contraction of the light cone reduced circuit, or  MPS dynamics with an analytically known  bond dimension that entails no truncation.
In Fig.~\ref{fig:hwZNE}(c), we report results of the weight-17 operator $X_{\{8\}} Y_{\{8\}} Z_{\{1\}} = \prod_{i}e^{-i \theta_h {X_i}/2} U(\pi/2)^5 Z_{58} \left[ U^\dagger(\pi/2) \right]^5 \prod_{i}e^{i \theta_h {X_i}/2}$ measured at circuit depth 5 with an additional layer of $\prod_{i}e^{-i \theta_h {X_i}/2}$ gates at the end.
The expectation value of this operator is entirely equivalent to the expectation value of $U(\pi/2)^6 Z_{58} \left[U^\dagger(\pi/2)\right]^6$ measured at circuit depth 6.
We find good agreement between experiment and Heisenberg results and additionally find the known answers at the two Clifford angles.
Given the convergence in bond dimension and reasonable circuit fidelities shown in Appendix~\ref{sec:numerics}, this demonstrates the applicability of ZNE beyond the exactly verifiable regime.

We next turn to single-site magnetization $\langle Z_{62} \rangle$ at depth 20, already shown in Fig.~\ref{fig:Z62}. 
Before discussing the results, let us describe the 6 different numerical methods: (1) Heisenberg MPO evolution described in Sec.~\ref{sec:heisenberg}, (2) pure state MPS evolution shown in~\cite{utility}, (3) extrapolation of pure state MPS results with respect to the estimated fidelity $F_D \to 1$ (see Appendix~\ref{sec:numerics}), (4) the recently developed heavy-hexagon belief propagation TNS (BP-TNS) method~\cite{tindall2023efficient}, (5)  Clifford Perturbation Theory~\cite{begušić2023simulating,begušić2023fast}, and (6) simulations of a 31 qubit subset of the device~\cite{kechedzhi2023effective}.
Methods (1), (2), and (4) are run at a fixed bond dimension $\chi$, and confidence in the results stems from apparent convergence in bond dimension, truncation error, and for (1) and (2), the circuit fidelity. Method (5) was run at order $K=10$.

We find that all numerical methods agree broadly with the experimental ZNE results, yet differences are present across the range of single qubit angles $\theta_h$. 
We find very good agreement between all methods and experiment for $\theta_h \leq \pi/8$. 
For $\theta_h > \pi/8$, the experimental error bars fall in between the results from Heisenberg MPO evolution and BP-TNS evolution.
As the MPO evolution is exact as $\chi \to \infty$, a natural improvement would be to extrapolate the MPO results with respect to the estimated circuit fidelity, as we did for pure state MPS. However, we often find the results are not monotonic in $\chi$, as shown in Appendix~\ref{sec:numerics}, making such extrapolation difficult. 
Further improvements in both the classical methods and experiment will be necessary to resolve the remaining discrepancy.

Why do MPO calculations prove sufficient at this scale, despite the naive scaling $\chi \sim k^{D^2}$? To investigate we characterize the operator growth in the Kicked Ising circuit, which governs the difficulty of MPO simulations. 
We first consider the operator entanglement entropy (OEE) of the Heisenberg-evolved $Z_{62}(D)$.
The OEE is defined by interpreting an operator as a wavefunction in a doubled Hilbert space,~\cite{prosen2007operator} and computing the resulting entanglement entropy of a bipartition; we focus on the largest OEE across all bipartitions of the 1D ordering. 
Focusing on $\theta_h = 0.7$, we observe a quadratic growth with circuit depth, $\sqrt{O_{EE}} \sim 0.11 D$ (the plateau at large times is a numerical artifact of finite-$\chi$ truncation; see Fig.~\ref{fig:whyWorks}(a)). This scaling is consistent with an operator growing over a disk of area $\propto D^2$, but with a  prefactor far below the maximal amount allowed by the lightcone (e.g., $O_{EE} \leq 0.7 D^2$ at $D=7$). 
A priori this may either because the ``butterfly velocity'' $v_B$, which governs the rate of operator spread, is less than than that of the lightcone, or because the nearby Clifford point delays the growth in operator entanglement.

In Fig.~\ref{fig:whyWorks}(c), we plot the out of time ordered correlator (OTOC) defined as $C(D, x) = \langle Z_x Z_{62}(D) Z_x Z_{62}(D)\rangle $, which is straightforward to evaluate once $Z_{62}(D)$ is obtained in MPO form\footnote{We simply evaluate the expectation value of the doubled operator $Z \otimes Z$ using the vectorized MPO as an MPS.}. 
At $D=7$, we see that the OTOC signal is confined well within the lightcone, indicating $v_B$ quite a bit below 1. 
This was also recently observed in Ref.~\cite{kechedzhi2023effective}, where it was  exploited to efficiently simulate the Kicked Ising circuits with reduced qubit count. 

\section{Comparison of operator methods with BP-TNS}
\label{sec:BPTNS}
Belief propagation tensor network states (BP-TNS) \cite{sahu2022efficient, tindall2023efficient} on the heavy-hexagon lattice were recently used to simulate the Kicked Ising experiments and were found to match the exact results where available and provide good agreement with the ZNE results more generally~\cite{tindall2023efficient}. The BP-TNS method shows great promise for approximating the local observables of moderate-depth dynamics, so we take a moment to consider some strengths and weaknesses relative to the operator based approaches. 

BP-TNS is two-dimensional (2D) tensor network (PEPs) ansatz in which the ``environment'' of each region (an input for the calculation of local expectations values) is approximated by assuming it takes the form of an uncorrelated tensor-product across ancilla bonds. Under this approximation - which is uncontrolled -  the environment can be self-consistently calculated in a procedure analogous to statistical ``belief propagation''~\cite{sahu2022efficient}.
In contrast to 1D MPS / MPO methods, BP-TNS need not reproduce the exact results even as the bond dimension $\chi \to \infty$, a point we will return to.

\begin{figure}
    \centering
    \includegraphics{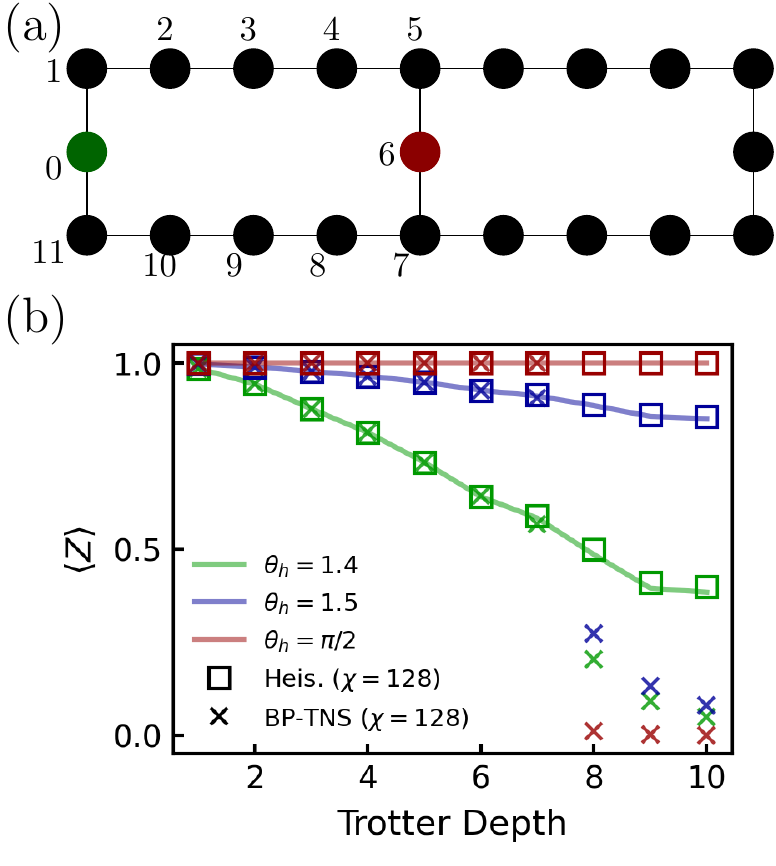}
    \caption{Expectation value of the depth-$D$ Clifford stabilizer $\mathcal{S}_D = U(\pi/2)^D Z_i \left[U^\dagger(\pi/2) \right]^D$ with respect to the depth-$D$ circuit $U(\pi/2)^D$.
    BP-TNS calculations are conducted at $\chi = 2^7 = 128$. We see that once $\chi < 2^D$ there is a catastrophic loss in accuracy, as expected for any pure-state TNS method. 
    Heisenberg simulations remain highly accurate due to the proximity to the Clifford point, though still suffer from an exponential blowup more generally.}
    \label{fig:echo}
\end{figure}

Like other 2D TNS methods, BP-TNS naturally adapts to the 2D geometry of the heavy hexagon lattice, avoiding the need of MPS/MPO algorithms to turn short-range 2D interactions into long-range 1D interactions. Additionally, approximating the environments as uncorrelated reduces the computational cost compared to more expressive 2D tensor network algorithms (e.g. PEPs / isoTNS)~\cite{verstraete2004renormalization, Zaletel_2020}, allowing BP-TNS to scale to much larger bond dimensions ($\chi\sim200$) than is possible with standard 2D TNS ($\chi\leq 12$).
This is crucial for the kicked TFI circuits near the $\theta_h=\pi/2$ Clifford point, where the required bond dimension for an accurate 2D TNS representation scales as $2^D$.
In particular, at $\theta_h=\pi/2$ the  entanglement spectrum on any bond of the network, given by the diagonal $\Lambda$ matrices of BP-TNS, has rank $2^D$ and is \textit{exactly} flat.
Thus any truncation of the bond dimension will immediately lead to noticeable errors in expectation values.

The effect of BP-TNS bond truncation can be seen in Figure~\ref{fig:echo}, where we use BP-TNS to measure a high-weight stabilizer $\mathcal{S}_D \equiv U(\pi/2) ^D Z_i \left [U^\dagger(\pi/2) \right]^D$ of the $\theta_h = \pi/2$ circuit as a function of depth $D$. In order to compare with brute-force exact results we consider the ``two-hexagon'' geometry with 21 qubits, shown in Figure~\ref{fig:echo}(a), which retains the salient features of the 127 qubit geometry.
Using the ``extended time-evolution'' method proposed in Ref.~\cite{tindall2023efficient}, we measure $\mathcal{S}_D$ by first evolving a circuit forward $D$ steps with single-qubit angle $\theta_h$ and then backward $D$ steps with angle $\pi/2$: $\ket{\psi} = \left [U^\dagger(\pi/2) \right]^D U(\theta_h)^D \ket{\uparrow}^N$.
This allows for high-weight $\mathcal{S}_D$ to be measured as a single-site observable: $\braket{\uparrow | \left[ U^\dagger(\theta_h) \right]^{D}  \mathcal{S}_D U(\theta_h)^D  | \uparrow} =  \braket{\psi | Z_i| \psi} $, where $i$ is shown as a red qubit in Figure~\ref{fig:echo}(a).
We consider up to depth $D= 10$ with fixed bond dimension $\chi = 128$ for both BP-TNS and Heisenberg evolution.
In Figure~\ref{fig:echo}(b), we demonstrate that as one approaches the Clifford point $\theta_h=\pi/2$, using a BP-TNS bond dimension less than $2^{D}$ leads to a catastrophic reduction in the expectation value.
The difficulties of BP-TNS for large $\theta_h$  occur for the same reason MPS and isoTNS pure-state simulations struggled in the original simulations of Ref.~\cite{utility}: the evolved state is highly entangled.
Heisenberg evolution, on the other hand, reproduce the dynamics of the $\pi/2$ Clifford point, regardless of circuit depth $D$, with bond dimension $\chi=1$ (or CPT order $K=0$) as the evolved operator is a single Pauli string.
Of course away from the Clifford points these methods also suffer from an eventual exponential blowup, but the difficulty is pushed out to  higher depths. 

\begin{figure}[tb]
    \centering
    \includegraphics{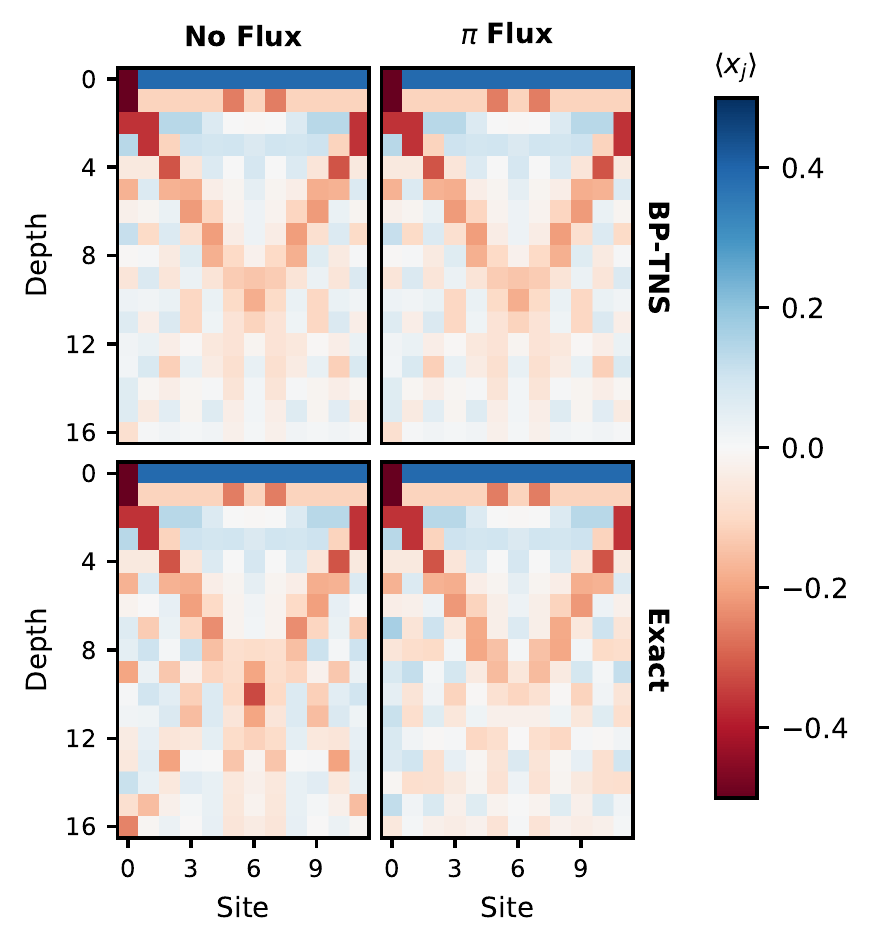}
    \caption{Kicked-Ising ($\theta_J = -\pi/4,\, \theta_h = \pi/2$) realization of a many-body double slit experiment. An initial $x$-polarized state is prepared with a spin flip on site 0 of a double-hexagon geometry (Fig.\ref{fig:echo}a), which then propagates under kicked-Ising dynamics as revealed by the site-resolved magnetization. In the exact result (bottom row), the magnetization density exhibits constructive / deconstructive interference on site 6, $D\sim10$ depending on whether $\phi = 0 / \pi$ flux threads the left hexagon. 
    Within the BP-TNS approximation (top row), there is no sign of interference: the result is entirely independent of the flux through the hexagon. 
    This is because the BP-TNS approximation cannot capture the loop-like entanglement required by the double-slit experiment. 
    }
    \label{fig:double_slit}
\end{figure}

\emph{The effect of loop-like entanglement.} A more interesting aspect of BP-TNS arises from its approximate treatment of the TNS ``environment.''
Even if $\chi = 2^D$, the BP-TNS environment approximation becomes inexact once the lightcone of any site encompasses a loop within the lattice. As pointed out in Ref.~\cite{tindall2023efficient}, BP-TNS is thus well suited to the heavy-hex lattice, whose shortest loop has length 12. For the kicked TFI circuit $U(\theta_h)$,  BP-TNS is then exact up to depth $D=6$ if bond dimension $2^6 = 64$ is used, which  happens to be depth of the circuits amenable to brute-force calculation where verification was conducted. 
Beyond  $D=6$, the BP-TNS makes uncontrolled approximations, and it is an interesting question whether the Kicked Ising circuit ever features the ``loop-like'' correlations which evades the BP-TNS approximation.
At $D=20$, it is at least quantitatively suggestive that near the $\theta_h \sim \pi/4$ point, there is a $20\%$ discrepancy in $\langle Z_{62} \rangle$ between BP-TNS and the CPT / Heisenberg methods \emph{even once extrapolating} the BP-TNS $\chi \to \infty$. This is even more clear near $\theta_h  = 3 \pi / 8$, where ZNE, CPT, and Heisenberg all give $\langle Z_{62} \rangle \sim 0$, while BP-TNS gives $\langle Z_{62} \rangle \sim 0.05$. 
CPT and Heisenberg methods are expected to be quite accurate here. 

One fanciful way to illustrate the generation of loop-like entanglement in kicked-Ising dynamics is to conduct a many-body version of the two-slit experiment. 
To do so we consider a modified circuit in which the $ZZ$-gates are made non-Clifford: $V = U(\theta_J=-\pi/4,\, \theta_h=\pi/2) = R_{ZZ}(-\pi/4) R_X(\pi/2)$.
We then consider a state polarized along $X=1$, flip a spin at site $i$, and measure the change in $X$-magnetization at site $j$:
\begin{align}
C_{ij}(D) = \bra{\rightarrow^N} Z_i \left[V^{\dagger}\right]^D X_j V^D Z_i \ket{\rightarrow^N}.
\end{align}
The dynamics cause the spin flip to propagate from $i \to j$, albeit on top of a rapidly thermalizing background. 
Choosing $i / j$ to lie on the green / red sites of Fig.~\ref{fig:echo}(a), the dynamics of the spin flip effectively realize the two-slit experiment as the spin flip enters into a superposition of the top and bottom arms of the hexagon. 
In Fig.\ref{fig:double_slit}, we show the dynamics of the $X$-magnetization in the left 12-site hexagon for both the exact and BP-TNS approximation ($\chi = 128$, re-gauging with 15 ``message passing'' iterations). 
In the exact result, we see a wave front which propagates and constructively interferes when the top and bottom fronts collide on the ``red'' site. 
BP-TNS reproduce the exact results beautifully before the collision but fail to capture the interference phenomena as the fronts overlap. 
This is because a spin-flip in such a superposition is exactly the sort of entanglement neglected in the BP-TNS approximation: the influence of the top and bottom arms on the ``red'' site is assumed to factorize.  
This failure can be quantified by threading $\pi$-flux (e.g. flipping the sign of one $ZZ$-bond) through the hexagon and repeating  the experiment. 
In the exact simulation, we see the wavefronts now interfere \emph{destructively}; BP-TNS, in contrast, gives identical results for both fluxes. 

We do not expect such a simple picture to apply to deeper circuits; the double-split example required a coherence length comparable to the length of the minimal loop. As the loop develops more connections to the surrounding geometry, or as the effective energy density increases due to the driving, we expect additional many-body degrees of freedom will obtain ``which-path'' information which destroys the interference, and the BP-TNS approximation may become more accurate. 

\section{Circuit extensions}

Having highlighted some strengths and weaknesses of both pure state and operator evolution methods, we now propose extensions to experiments that may lead to more challenging simulations, both classically and for ZNE.

\emph{Non-Clifford two qubit gates.} Reducing the two-qubit angle away from the Clifford point, say  $\theta_J \to -\pi/4$ will increase the difficulty of the CPT and MPO-Heisenberg methods as it increases the rate at which new Pauli strings are generated. 
In Fig.~\ref{fig:whyWorks}(a,b), we compute the $O_{EE}$ and MPO fidelity at $\theta_J = -\pi/4, \theta_h = 0.7$.
We see $\sqrt{O_{EE}} \sim 0.18 D$, almost 2x the rate of the Clifford 2Q case, and correspondingly the MPO fidelity drops much faster with gate depth. 
Note, however, that a non-Clifford two-qubit gate may make pure-state simulations such as BP-TNS easier, as such a gate is less entangling than one with $\theta_J=-\pi/2$.

\emph{Non-commuting two qubit gates.} 
$R_{ZZ}$ is composed of three layers of disjoint $ZZ$ gates. While naively each layer could advance the lightcone, because they commute, the lightcone grows by only one site in each direction per $R_{ZZ}$.
The simplification is obstructed if the two-qubit gates are non-commuting; or equivalently, by sprinkling  the $R_X$ gates amongst the three layers.
In Fig.~\ref{fig:whyWorks}(a,b), we compute the $O_{EE}$ and fidelity at $\theta_J = -\pi/2, \theta_h = 0.7$, performing a round of $R_X$ gates after each layer of $ZZ$ gates, thus tripling the number of single qubit gates.
In addition to increasing the size of the lightcone, we find that the OEE grows much more rapidly than for the commuting kicked TFI models considered ($\sqrt{O_{EE}} \sim 0.33 D$), and the plateau imposed by finite bond dimension is much higher than for the commuting version also with $\theta_J=-\pi/2$.
However, the region with non-unital OTOC is still a fraction of the circuit lightcone, indicating $v_B < 1$ but higher than the commuting case. Experimentally, this modification has the advantage of an identical two-qubit gate count.

\begin{figure}[tb]
    \centering
    \includegraphics{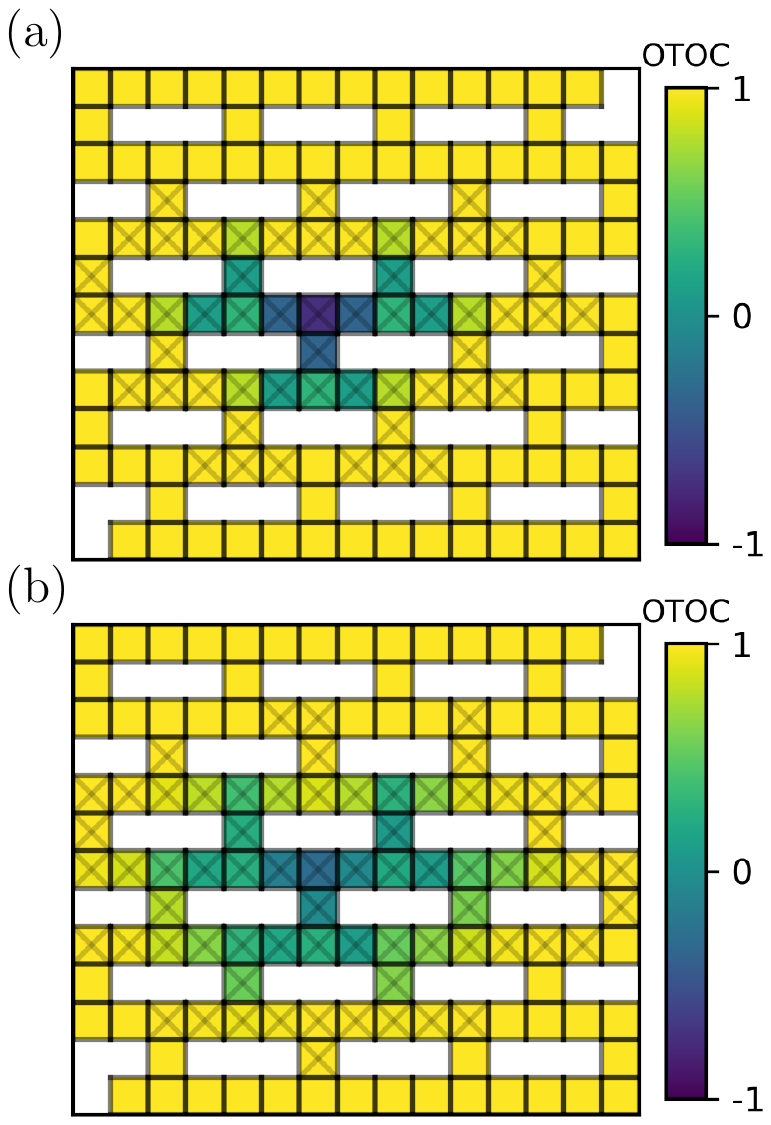}
    \caption{OTOC growth for Kicked TFI variants. The lightcone at depth $D=7$ is depicted by the crosses. (a) $C(7, x)$ for the $(\theta_J, \theta_h) = (-\pi/4, 0.7)$ model. Compared to the version with $\theta_J=-\pi/2$, shown in Fig.~\ref{fig:whyWorks}(c), we see that the OTOC grows both in size and spatial extent by depth 7. (b) $C(4, x)$ for the $(\theta_J, \theta_h) = (-\pi/2, 0.7)$ non-commuting model introduced in the text. We see the lightcone at depth $D=4$ is larger (69 sites) compared to that for commuting models (54 sites) at a greater depth and that the OTOC has spread more than it has in other models.}
    \label{fig:nonClifford}
\end{figure}

Of course at fixed $\theta_h$, the final signal $\langle Z \rangle$ depends on these modifications as well, and for ZNE to be performant it must be measurable.
In the non-trotterized dynamics ($\theta_h, \theta_J \to 0$ holding $\theta_h/ \theta_J$ fixed), we expect a continuous phase transition tuned by $\theta_h/ \theta_J$, corresponding to finite-$T$ equilibrium symmetry breaking of the Ising model.
Even at finite Trotter step, the moderate-time dynamics will be affected by the transition before ultimate ergodicity, making this an interesting parameter regime for comparison of ZNE and classical approaches. 

\emph{Echo experiments.} 
From the perspective of benchmarking ZNE, another future direction is to measure the  analog of high-weight operators designed to ensure a large signal. 
Away from the Clifford point the high-weight operators are complex, so can instead be measured according to
\begin{align}
Z_i(D|\theta, \theta') \equiv & \bra{\textrm{CPS}} \left[U^\dagger(\theta')\right]^{D} U(\theta)^D  \nonumber \\
& Z_i \left[U^\dagger(\theta)\right]^{D} U(\theta')^D \ket{\textrm{CPS}}
\end{align}
Such an experiment is essentially a Loschmidt echo, but restricting the return probability to a single site rather than the global fidelity.

\section{Conclusion}

We have simulated the circuits considered in Ref.~\cite{utility}, which we previously studied with pure state methods, via  matrix product compression Heisenberg evolution of the operators. 
We match the exact answer where available and find general agreement with ZNE experimental results.
For the largest circuit depths considered, we find 10 - 20\% disagreements  between other recently reported classical methods ~\cite{tindall2023efficient, begušić2023fast, kechedzhi2023effective}.
The classical uncertainty appears to be within the bounds of the ZNE uncertainty, and likely can be further reduced with additional resources. 
The Heisenberg approach reveals a relatively slow growth of operators in the studied parameter regime, and modifications to the circuits which would make similar calculations more difficult are discussed.

\begin{acknowledgments}
    SA and MZ were supported by the U.S. Department of Energy, Office of Science, Basic Energy Sciences, under Early Career Award No. DE-SC0022716. We thank Y. Kim and A. Eddins for insightful experimental data and simulations. We thank S. Bravyi for suggestions on additional simulations. We thank J. Tindall, M. Fishmann, M. Stoudenmire, K. Siva, T. Soejima, and S. Garratt for helpful discussions.
    Computing resources were provided by National Energy Research Scientific Computing Center (NERSC), a U.S. Department of Energy Office of Science User Facility located at Lawrence Berkeley National Laboratory, operated under Contract No. DE-AC02-05CH11231 using NERSC Award No. BES-ERCAP0020043. 

\end{acknowledgments}

\appendix
\section{Additional numerics}
\label{sec:numerics}
Here we present additional numerics for the simulations discussed in the main text.
\subsection{Exactly verifiable operators}
\begin{figure}[tb]
    \centering
    \includegraphics{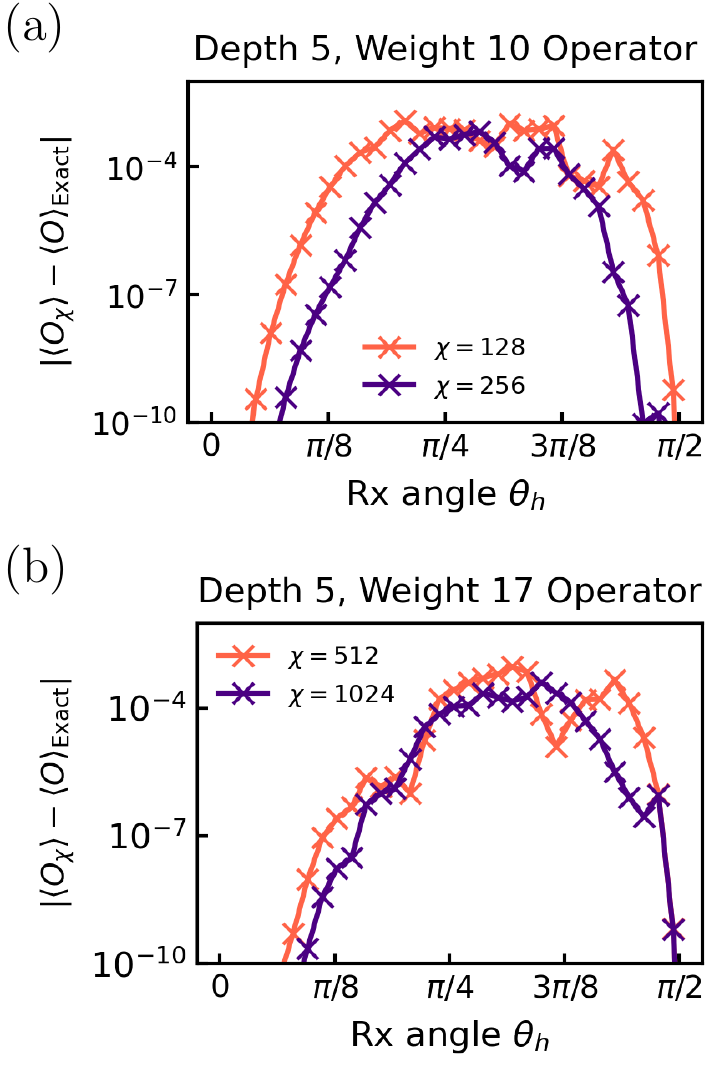}
    \caption{Error for Heisenberg simulation of the (a) weight-10 and (b) weight-17 operators where the exact solution is available. We generally find accurate quantitative agreement with the exact result across a wide range of $\theta_h$, building confidence when moving to more difficult circuits.}
    \label{fig:convergence}
\end{figure}

We begin in Fig.~\ref{fig:convergence} with the error in simulations of the exactly solvable weight-10 and weight-17 observables shown in Fig.~\ref{fig:hwZNE}(a,b).
We compare results with Heisenberg evolution to the exact answer and find good agreement across a range of $\theta_h$ values. 
Note that $\chi=384$ is sufficient to obtain the exact answer for the weight-10 operator, as no truncation occurs during the depth-5 simulations.
For the weight-17 operator, we find that increasing the bond dimension does not always reduce the error, as is not uncommon for time evolution which is not variational.
Individual simulations for the weight-10 (17) operator take less than 2 (90) minutes at the maximum bond dimensions considered.

\subsection{MPO Circuit Fidelity}
Before moving on to more difficult circuits, let us introduce a metric to bound the accuracy of matrix-product operator based time evolution. 
Recall that at each Trotter step, i.e. each application of the 13 MPOs that apply a layer of two-qubit gates on every bond on the lattice, the bond dimension $\chi$ grows and must be truncated via variational compression. 
Upon compressing the evolved an enlarged MPO $\tilde{O}_t = U_r^\dagger O_{t-1} U_r \to O_t$ back to fixed $\chi$, a truncation error $\epsilon_t   = |\tilde{O}_t - O_t|$ is incurred. 
Here we use the Frobenius norm, scaled so that all Pauli strings have unit norm. 
We can then upper-bound the total error after $D$ steps by $\epsilon(D, \chi)  \leq \sqrt{ \sum_t \epsilon_t^2}$. 
We record this as a fidelity, $f_t = |\braket{O_t| \tilde{O}_t}|^2$,  so that $F_D \equiv \prod_{t=1}^D f_t \leq e^{-\sum_{t=1}^D \epsilon_t^2} \leq e^{-\epsilon^2(D, \chi)}$.
As $\chi\to\infty$, $F_D\to 1$ as each step of the evolution will be done exactly.
We use the fidelity to both extrapolate results from finite bond dimension and qualitatively gauge the confidence in our simulation.
An equivalent estimate of the final fidelity is also applicable in the context of MPS pure states, as discussed for example in Ref.\cite{zhou2020limits}.

\subsection{Modified Weight-17 Operator}
\begin{figure}[tb]
    \centering
    \includegraphics{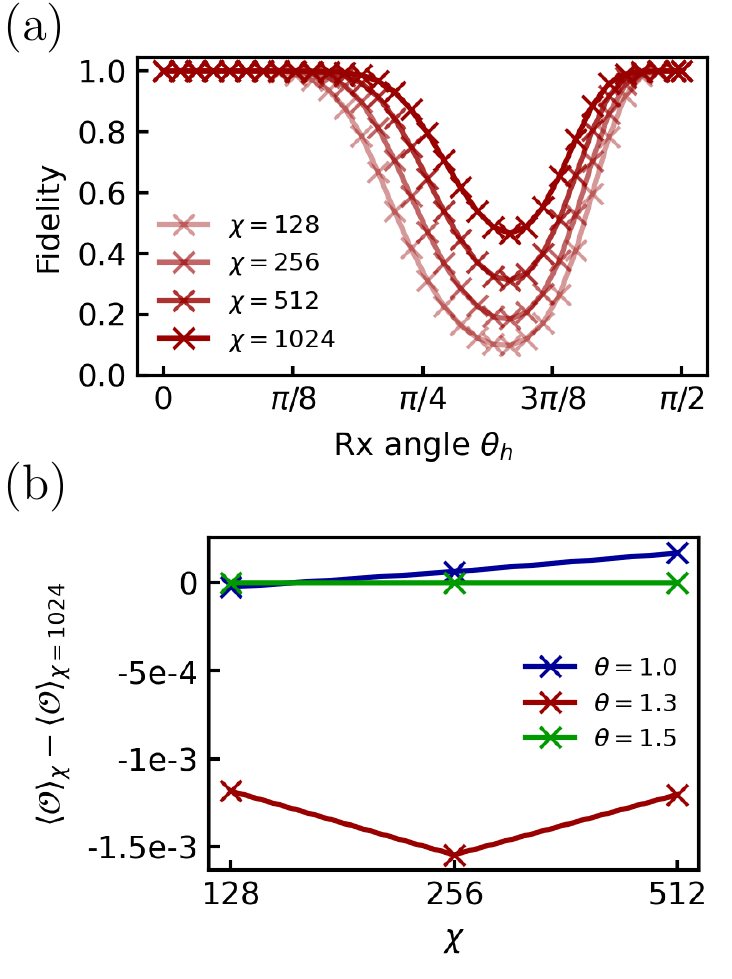}
    \caption{(a) Fidelity as a function of single qubit angle $\theta_h$ for different $\chi$ for Heisenberg simulation. Near the Clifford points, the simulations have fidelity $1$ with small bond dimension, but there is a significant dip in the intermediate regime. (b) Convergence of $\langle \mathcal{O} \rangle$ as a function of $\chi$. As the exact answer is not available, we look at differences from the value at $\chi=1024$, the largest bond dimension considered. Non-monotonic behavior in $\chi$ is observed.}
    \label{fig:mwt17_convergence}
\end{figure}

Next we turn to the weight-17 operator for the modified circuit: 5 layers of the circuit defined by Eq.~\ref{eq:utilityTFI}, followed by an additional layer of $R_X$ rotation gates.
The depth-5 stabilizer of this circuit is equivalent to the depth-6 stabilizer of the original circuit and was chosen to access this larger operator with the same circuit resources.
In Fig.~\ref{fig:mwt17_convergence}(a), we show the circuit fidelity $F_D$ as a function of $\theta_h$ for several bond dimenions.
We see that at the two Clifford points $\theta_h=0,\, \pi/2$ that the fidelity is exactly 1, as a bond dimension of $\chi=1$ is sufficient to track the evolution of a Pauli string.
We see that the fidelity increases demonstrably with increasing chi, which as been argued to indicated that the exact bond dimension for this circuit isn't exponentially far removed from what is achievable in simulations~\cite{zhou2020limits}.
In Fig.~\ref{fig:mwt17_convergence}(b), we investigate the convergence of the expectation values at different $\theta_h$ as a function of bond dimension.
Note that the expectation value is not changing monotonically with bond dimension.

\subsection{$\langle Z_{62} \rangle$ - MPS Extrapolation}
\begin{figure*}[tb]
    \centering
    \includegraphics{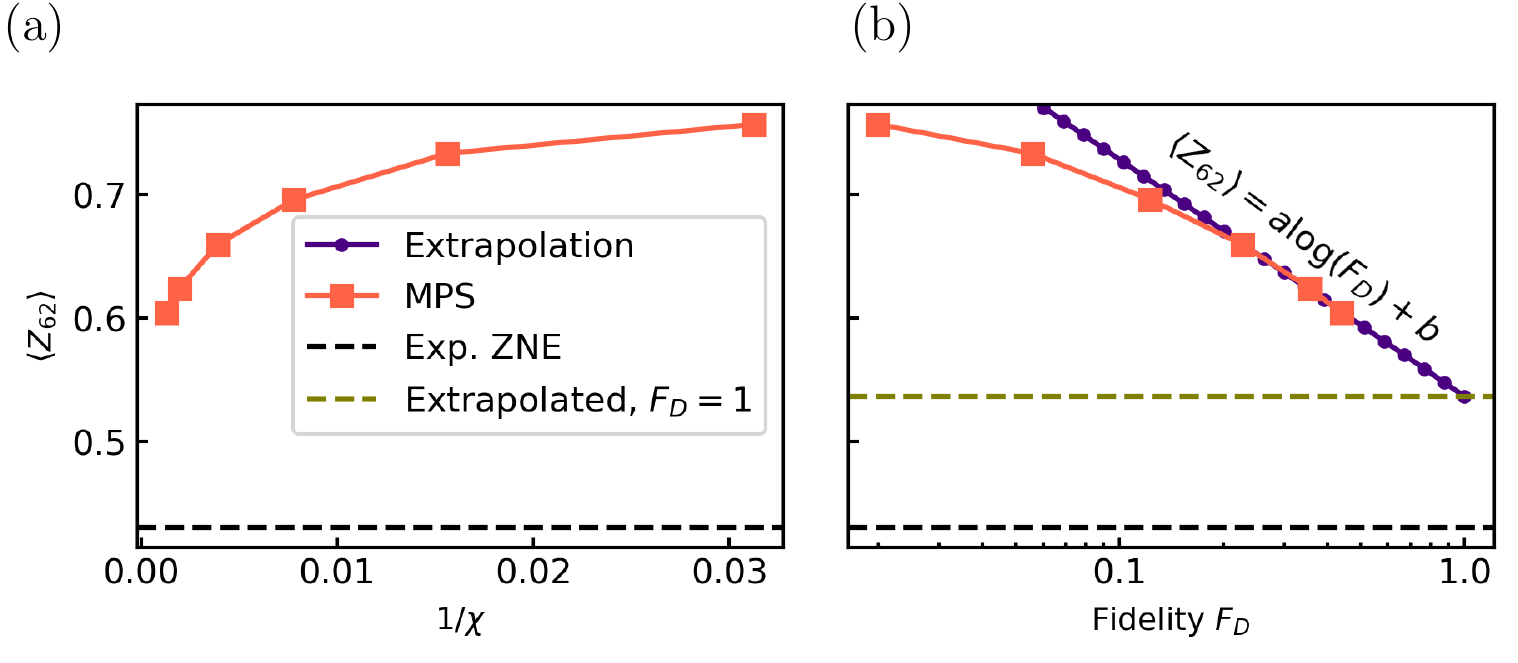}
    \caption{Extrapolation of $\langle Z_{62} \rangle$ using circuit fidelity. (a) Expectation value as a function of $1/\chi$ does not display a clear power law for extrapolation. (b) Extrapolation performed at $\theta_h=0.7$ with circuit fidelity, $\langle Z \rangle = a \cdot \log(F_D) + b$.}
    \label{fig:MPSextrapolation}
\end{figure*}

Finally we turn to the depth-20 circuits for $\langle Z_{62} \rangle$.
In Fig.~\ref{fig:MPSextrapolation} we demonstrate how we extrapolate the pure state, Schr\"{o}dinger picture MPS simulations from finite bond dimension to $\chi=\infty$.
A clear trend in $\langle Z \rangle$ in $1/\chi$ is not seen in Fig.~\ref{fig:MPSextrapolation}(a), so we instead use our circuit fidelity.
We fit the results for the largest three bond dimensions as a linear function in $\log(F_D)$ and extrapolate to $F_D \to 1$.
As shown in Fig.~\ref{fig:Z62}, extrapolating MPS data to $F_D=1$ (achieved at $\chi=\infty$ where each step of the time evolution can be done exactly as no truncation is ever done) brings the results closer to the ZNE experiment as well as the other computational methods, but it is not sufficient to find agreement.

\subsection{$\langle Z_{62} \rangle$ - MPO Extrapolation}
\begin{figure*}[tb]
    \centering
    \includegraphics{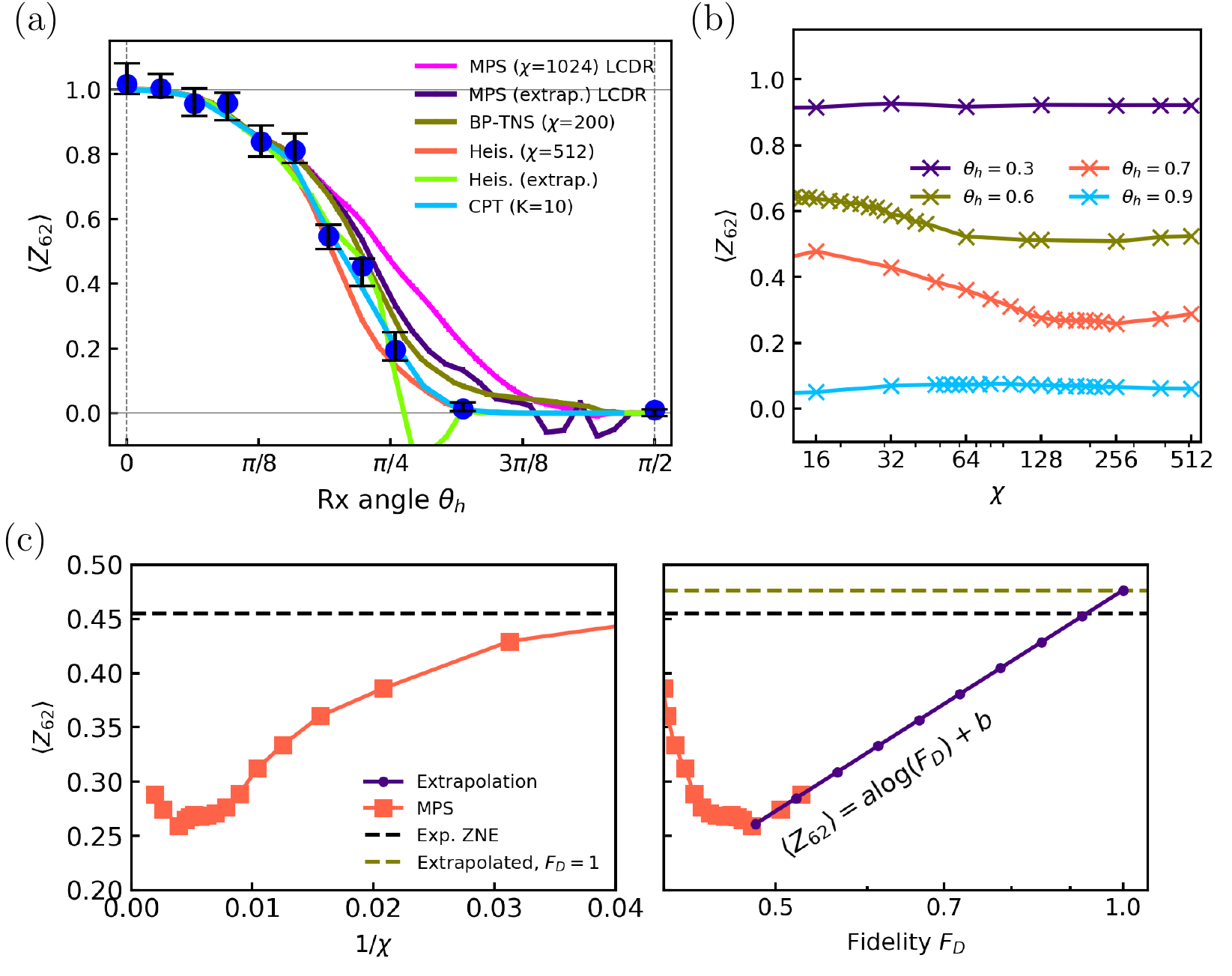}
    \caption{(a) Attempted extrapolation of Heisenberg results with circuit fidelity show behavior inconsistent with other classical methods and ZNE experimental results at intermediate $\theta_h$. (b) Examples of non-monotonic behavior found at various $\theta_h$. (c) Example of extrapolation performed at $\theta_h=0.7$.}
    \label{fig:Heisextrapolation}
\end{figure*}

A natural next step is to try extrapolation of the Heisenberg results in circuit fidelity. We show the results of this in Fig.~\ref{fig:Heisextrapolation}(a), where the extrapolation for $\theta_h=0.7$ is shown in Fig.~\ref{fig:Heisextrapolation}(c).
Note that unlike for the MPS simulations, the expectation value is often non-monotonic in $\chi$ (or equivalently $F_D$), which can be seen in Fig.~\ref{fig:Heisextrapolation}(b), which makes extrapolation less justified.
To demonstrate, we extrapolate the same way as with MPS, using the three largest bond dimension results.
We see that the extrapolated results are reasonable (and lie amongst the other classical methods and ZNE experimental results) for $\theta_h \leq \pi/4$, but for $\pi/4 < \theta_h \leq 3\pi/8$ are generally inconsistent with other methods.
The expectation value in this region is decreasing with $\chi$, which leads to the negative extrapolated value.
If trends at smaller $\theta_h$ are to continue, the expectation value will begin increasing with $\chi$ after some point, but the necessary $\chi$ to see this may be exponentially large~\cite{zhou2020limits}.

\bibliographystyle{apsrev4-1} 
\bibliography{refs} 

\end{document}